\documentclass{aa}

\usepackage{graphicx}
\usepackage{float}
\usepackage{color}
\usepackage{amsmath}
\usepackage{multirow}
\usepackage[colorlinks=true, linkcolor=blue, citecolor=blue, urlcolor=blue]{hyperref}

\usepackage{natbib}
\bibpunct{(}{)}{;}{a}{}{,}

\usepackage{txfonts,textcomp}

\makeatletter
\let\linenumbers\@gobble
\let\endlinenumbers\@empty
\makeatother

\begin{document}

    \title{Tentative Detection of the Glycine Isomer Glycolamide in Hot Molecular Core
                      \thanks{Corresponding authors: \\
                      Qian Gou, \email{qian.gou@cqu.edu.cn}\\
                      Xuefang Xu, \email{xuefang\_xu@cqu.edu.cn}}}

   \subtitle{}

   \author{Chunguo Duan\inst{1,2},
          Fengwei Xu\inst{3},
          Qian Gou\inst{1,2}, 
          Xuefang Xu\inst{1,2},    
          Donghui Quan\inst{4},               
          Laurent Pagani\inst{5},                       
          Xi Chen\inst{6},       
          Jun Kang\inst{1,2},
          Jiaxin Du\inst{1,2}
          }

   \institute{
            \inst{1} School of Chemistry and Chemical Engineering, Chongqing University, Daxuecheng South Rd. 55, Chongqing 401331, People’s Republic of China\\
            \inst{2} Chongqing Key Laboratory of Chemical Theory and Mechanism, Chongqing University, Daxuecheng South Rd. 55, Chongqing 401331, People’s Republic of China\\
            \inst{3} Max Planck Institute for Astronomy, Königstuhl 17, Heidelberg 69117, Germany\\
            \inst{4} Department of Physics, Xi’an Jiaotong-Liverpool University, Ren’ai Road 111, Suzhou 215123, People’s Republic of China\\
            \inst{5} LUX, Observatoire de Paris, PSL Research University, CNRS, Sorbonne Universités, Paris 75014, France\\                                      
            \inst{6} Center for Astrophysics, GuangZhou University, Guangzhou 510006, People’s Republic of China\\               
           }

  \abstract 
   {Understanding whether prebiotic molecules can endure and reform through the energetic stages of star formation is essential for tracing the continuity of interstellar chemistry toward life. Glycolamide, an isomer of glycine, was recently detected in the molecular cloud G+0.693-0.027 \citep{2023ApJ...953L..20R}. However, establishing its presence in warm, high-density environments is crucial to evaluate the chemical continuity of amides. Here we report the tentative detection of glycolamide in a hot molecular core, G358.93-0.03 MM1, using ALMA 1 mm observations. Seven unblended or only mildly blended emission lines were identified, yielding an abundance of (1.7$\pm$0.2)$\times 10^{-10}$ relative to H$_{2}$. The comparable formamide/glycolamide and acetamide/glycolamide abundance ratios in both sources suggest a chemically connected amide network across different environments. These results demonstrate that amides can persist and chemically evolve during massive star formation, tracing the chemical continuity from interstellar to protostellar environments.}

   \keywords{ISM: molecules -- ISM: abundances -- Astrochemistry}

\titlerunning{Glycolamide} \authorrunning{Duan et al.}\maketitle

\section{Introduction} \label{sec:sec1}

The emergence of life on Earth dates back to $\sim$3.8 billion years ago \citep{2018AsBio..18..343P}, yet the astrochemical origin of its fundamental molecular building blocks remains unsettled. Among the wide variety prebiotic molecules, amino acids are of particular interest because they constitute the monomeric units of proteins, which play pivotal catalytic and metabolic roles in living systems \citep{2004Sci...303.1151P, 2006PNAS..10312713W, wu2009amino}. Their identification in meteorites \citep{1983AdSpR...3i...5C, 2012cosp...39..264B}, samples returned from asteroid Ryugu \citep{2023NatCo..14.1482P}, and comet 67P/Churyumov-Gerasimenko \citep{2016SciA....2E0285A} implies that part of amino acid chemistry may trace back to interstellar or protosolar stages, motivating systematic searches for these species in the interstellar medium (ISM).

Glycine (NH$_{2}$CH$_{2}$COOH), the simplest amino acid, has long been the prime target for astronomical searches because it is thought to be among the easiest bio-monomers to assemble under interstellar conditions \citep{ruiz2014prebiotic}. Despite this expectation, all previous attempts have failed to yield a secure detection. Early claims toward several high-mass star-forming regions (HMSFR) were latter refuted \citep{2003ApJ...593..848K, 2005ApJ...619..914S}. Subsequent, more sensitive observations toward the HMSFR SgrB2 \citep{1996A&A...308..618C, 2007MNRAS.374..579J, 2007MNRAS.376.1201C}, Orion KL \citep{1996A&A...308..618C, 2007MNRAS.376.1201C}, the molecular cloud G+0.693-0.027 \citep{2020AsBio..20.1048J}, the low-mass protostar IRAS 16293-2422 \citep{2020AsBio..20.1048J, 2000A&A...362.1122C}, and the pre-stellar core L1544 \citep{2016ApJ...830L...6J} have also not confirmed its presence. This persistent non-detection has reshaped the field’s approach—from direct glycine searches toward indirect chemical diagnostics, focusing on its structural isomers and related species that may be more abundant or more readily observable.

Glycine belongs to the C$_{2}$H$_{5}$O$_{2}$N family, several members of which have been characterized spectroscopically, including methyl carbamate \citep[CH$_{3}$OC(O)NH$_{2}$;][]{Mar1999, BAKRI2002312, ILYUSHIN2006127}, the $syn$- and $anti$- isomers of glycolamide \citep[HOCH$_{2}$C(O)NH$_{2}$;][]{Maris04, 2020A&A...639A.135S}, and the $Z$- and $E$- isomers of acetohydroxamic acid \citep[CH$_{3}$C(O)NHOH;][]{2022A&A...666A.134S}. These spectroscopic studies have enabled dedicated astronomical searches for C$_{2}$H$_{5}$O$_{2}$N isomers toward a variety of interstellar sources, but most of them remain undetected \citep{2004sf2a.conf..493D, 2020A&A...639A.135S, 2021A&A...653A.129C, 2022A&A...666A.134S}. In this context, glycolamide in its $syn$-form ($syn$-HOCH$_{2}$C(O)NH$_{2}$, hereafter GA) is of particular interest because its hydroxyl and amide groups can form an intramolecular hydrogen bond, which stabilizes the structure and enhances its dipole moment, favoring its detection at millimeter wavelengths. GA was recently identified toward the G+0.693-0.027 molecular cloud with an abundance of $\sim$5.5×10$^{-11}$ relative to H$_{2}$ \citep{2023ApJ...953L..20R}. This detection provided the first observational evidence that a glycine structural isomer—a direct molecular analog of an amino acid—can form and persist under low-UV interstellar conditions.

Recent studies have suggested that GA may be of potential prebiotic interest, for instance, as a precursor to key prebiotic molecules: the carbon–oxygen bond cleavage yields acetamide (CH$_{3}$C(O)NH$_{2}$) and ethanolamine (HOCH$_{2}$CH$_{2}$NH$_{2}$)—a crucial molecular building block that serves as the head group in phosphatidylethanolamine, a constituent of both eukaryotic and bacterial membranes \citep{Bergantini25}. Furthermore, GA can participate in polycondensation reactions to form N-(2-amino-2-oxoethyl)-2-hydroxyacetamide, a derivative of the glycine dipeptide. This raises the possibility that GA might offer an alternative pathway for peptide formation in interstellar environments \citep{2011PCCP...13.7449S, 2023ApJ...953L..20R, Perrero25}.

Testing whether amides like GA endures as interstellar ices warm and reprocess in protostellar cores is essential, because star formation subjects molecules to intense heating and radiation \citep{2020ChJCP..33..668M}. If such amino-acid-related species can persist or reform under these conditions, it would demonstrate that amide chemistry is robust enough to survive the transition from interstellar clouds to nascent stellar systems. Hot molecular cores (HMCs), compact ($\leq$0.1 pc), dense (n$\geq$10$^{6}$ cm$^{-3}$), and warm (T$\geq$100 K) regions associated with massive protostars during the deeply embedded stage of star formation, provide ideal natural laboratories for such tests. During this stage, ice mantle sublimation, radical-radical reactions during warm-up, and subsequent gas-phase processing jointly drive the emergence of molecular complexity \citep{2021ApJ...922..206S, 2025ApJ...988...95D}. 

To explore whether amino-acid-related chemistry extends into these protostellar environments, we investigate the presence of GA in the HMC G358.93-0.03 MM1 (hereafter G358.93 MM1)—the brightest of eight submillimetre sources in the high-mass star-formation region G358.93-0.03 complex \citep[MM1-MM8;][]{2019ApJ...881L..39B}, has been chosen in this work. MM1 exhibits exceptionally line-rich spectra and a chemically diverse inventory, including prebiotic or amide-related molecules such as acetonitrile CH$_{3}$CN, cyanamide NH$_{2}$CN, glycolaldehyde CH$_{2}$OHCHO, methylamine CH$_{3}$NH$_{2}$, ethylene glycol (CH$_{2}$OH)$_{2}$, isocyanic acid HNCO, and formamide NH$_{2}$CHO \citep{2023Ap&SS.368...33M, 2023MNRAS.525.2229M, 2024RAA....24g5014M, 2024MNRAS.533.1143M}. This chemical richness makes G358.93 MM1 an ideal testbed to investigate GA and its potential linkage to other amides. 

Using ALMA 1 mm observations, we report the tentative detection of GA in an HMC. Its abundance relative to H$_{2}$ and to related formamide and acetamide have been derived to probe possible chemical connections and formation pathways. The structure of this paper is as follows. Section \ref{sec:sec2} outlines the observations and data reduction; Section \ref{sec:sec3} presents the spectral analysis; Section \ref{sec:sec4} discusses chemical implication; Section \ref{sec:sec5} summarizes the conclusions.

\section{Observation and data reduction} \label{sec:sec2}

\subsection{Observation}  \label{sec:sec2.1}

The high-mass star-formation region G358.93-0.03 was observed using the Atacama Large Millimeter/Submillimeter Array (ALMA) band-7 receivers (Project ID: 2019.1.00768.S; PI: Crystal Brogan). The observation of G358.93-0.03 was carried out on 2021 May 17, with a phase center of RA(J2000) = $17^{\rm h}43^{\rm m}10^{\rm s}$.000, Dec.(J2000) = $-29^\circ 51^\prime 46^{\prime \prime}$.000 with a total on-source integration time of 2963.52 s. A total of 47 antennas were used in the array, providing baseline lengths from of 14 m to 2517 m. J1550+0527, a standard quasar used for calibration, served as both the flux and bandpass calibrator, to determine the absolute flux scale and instrumental response. J1744-3116, located near the target field, was adopted as the phase calibrator to correct for short-term atmospheric and instrumental phase variations. The ALMA correlator was configured in dual-polarization mode , covering four spectral windows centered at 290.5-292.4, 292.5-294.3, 302.5-304.4, and 304.2-306.1 GHz. Each spectral window provided a resolution of 488.24 kHz, corresponding to a velocity resolution ($\delta$V) of $\sim$0.5 km s$^{-1}$.

\subsection{Data Reduction}  \label{sec:sec2.2}

Calibration of the visibility data and subsequent imaging were carried out using the Common Astronomy Software Application \citep[CASA, version 6.1.1.15;][]{Mcm07}. The standard ALMA pipeline was first applied for initial calibration, including corrections for bandpass, flux, and phase variations.

Because G358.93 MM1 is a line-rich source, we first imaged the four spectral windows without continuum subtraction. We then used the line-free channels obtained by the CASA pipeline task $findcont$ as an initial estimate for the continuum-fitting and iteratively exclude channels affected by line emission until no line-emission channels are found over 3$\sigma$. The remaining line-free channels were subsequently used for continuum fitting, while the line-emission channels were excluded from the continuum image construction. Three iterations of phase self-calibration were performed to improve phase stability, using solution intervals of ‘inf’, ‘30.25s’, and ‘int’, respectively. Each iteration employed a multi-term multi-frequency synthesize (MT-MFS) deconvolver with two Taylor terms and multiscale cleaning scales of 0, 5, 15, 50, and 150 pixels. Briggs weighting with a robust parameter of 0.5 was adopted throughout. After self-calibration, the continuum image quality improved substantially, yielding a synthetic beam of 0.15$^{\prime \prime}$×0.1$^{\prime \prime}$ (position angle -87.1$^\circ$) and a rms noise level of $\sim$20 uJy beam$^{-1}$. 

The final self-calibration solutions were applied to the original spectral-line data to correct phase errors across all channels. The continuum emission was then removed in the UV domain using the CASA task uvcontsub, where the line-free channels were fitted with a first-order polynomial. continuum-subtracted visibilities were imaged with the multiscale deconvolver, using the same weighting scheme and scales (0, 5, 15, and 50 pixels). The briggs weighting is used with a robust parameter of 0.5. The beam of each channels are restored into the common one for further wide-spectra analyses. The final line-cube sensitivity achieved was $\sim$0.5 K per 0.5 km s$^{-1}$ channel, consistent with the expected thermal noise after self-calibration.

\begin{figure*}[!htp]
        \centering
    \includegraphics[width=0.9\textwidth]{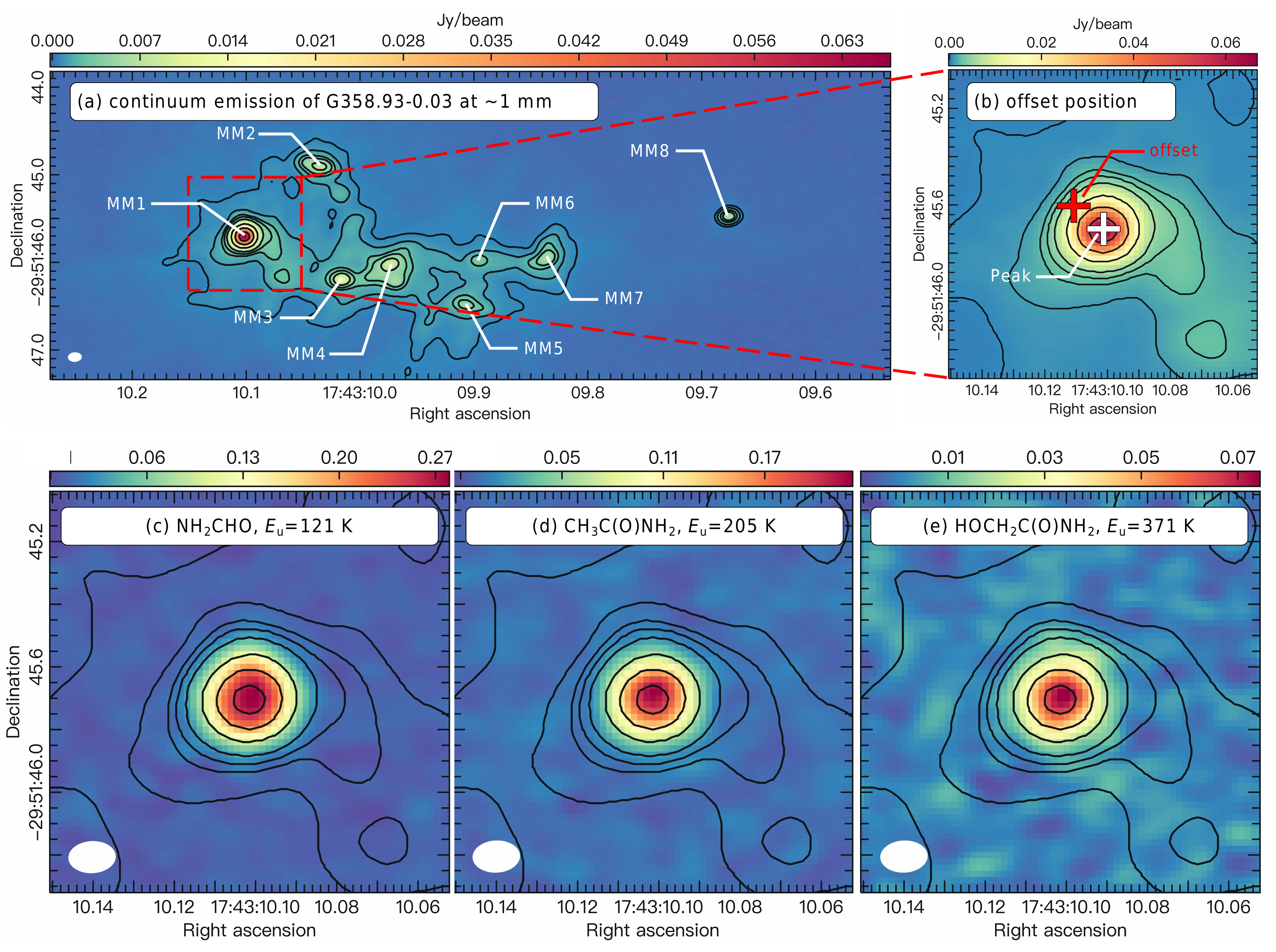}
        \caption{(a) 1 mm continuum image of G358.93-0.03. Contour levels correspond to (20, 40, 80, 120, 200, 500, 1000, 1700) $\times$ $\sigma$, where $\sigma$ is the rms noise level. The white ellipse in the lower-left corner indicates the synthesized beam 0.15$^{\prime \prime}$×0.1$^{\prime \prime}$ (position angle -87.1$^\circ$). (b) Enlarged view of MM1. The continuum peak is marked by a white cross, and the offset extraction position used for spectral analysis is marked by a red cross. (c-e) Integrated intensity (Moment-0) maps of NH$_{2}$CHO, CH$_{3}$C(O)NH$_{2}$, and $syn$-HOCH$_{2}$C(O)NH$_{2}$ toward G358.93 MM1, overlaid with 1 mm continuum emission contours from panel (a). The color scale unit is Jy beam$^{-1}$ km s$^{-1}$.
        \label{fig:1}} 
\end{figure*}

\begin{figure}[!htp]
    \centering
    \includegraphics[width=0.48\textwidth]{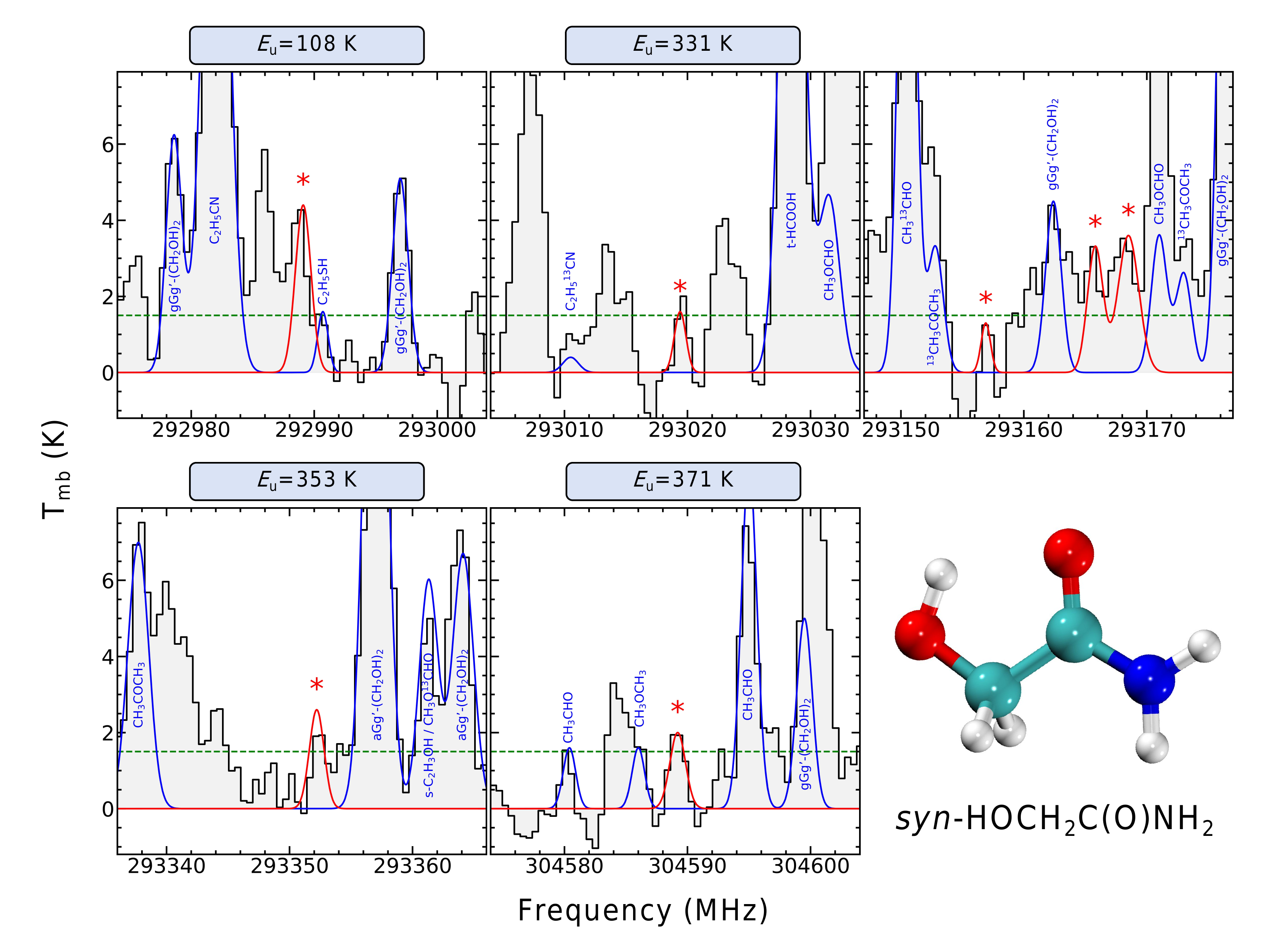}
    \caption{{Unblended or only slightly blended emission lines (red asterisks) of $syn$-HOCH$_{2}$C(O)NH$_{2}$ toward G358.93 MM1. Black lines show the observed spectra, while red lines represent the LTE model of $syn$-HOCH$_{2}$C(O)NH$_{2}$. Blue lines correspond to composite model including contributions from all other identified molecular species. Green dashed lines mark the 3$\sigma$ noise level. The upper energy level energies ($E_{\rm u}$) of the transitions are given above each panel.}}
    \label{fig:2}
\end{figure}

\section{Results} \label{sec:sec3}

\subsection{Spectroscopic inputs and local thermodynamic equilibrium modeling} \label{sec:sec3.1}

In this work, we focus on GA together with the chemically related amides formamide (NH$_{2}$CHO, hereafter FA) and acetamide (CH$_{3}$C(O)NH$_{2}$, hereafter AA). Spectroscopic data for these species were obtained from the Cologne Database for Molecular Spectroscopy\citep[CDMS\footnote{\url{https://cdms.astro.uni-koeln.de}};][]{Mul01, Mul05}. Table \ref{tab:A1} summarizes the corresponding database entries and literature sources used for the line parameters. 

The observed spectra were modeled under the local thermodynamical equilibrium (LTE) assumption using CLASS/Weeds \citep{Mar11} sets of Grenoble Image and Line Data Analysis Software (GILDAS\footnote{\url{http://www.iram.fr/IRAMFR/GILDAS}}). Five standard parameters were adopted in the model: source size ($\theta$), line width ($\Delta V$), source velocity ($V_{\rm LSR}$), excitation temperature ($T_{\rm ex}$), and column density ($N_{\rm t}$). To reduce parameter degeneracy, only $T_{\rm ex}$ and $N_{\rm t}$ were treated as free parameters, while the other quantities were fixed as follows: $\theta$ was set to the deconvolved continuum sizes, $\Delta V$ to the average of Gaussian fits for bright unblended lines, and $V_{\rm LSR}$ to -16.5 km s$^{-1}$ for G358.93 MM1 \citep{2019ApJ...881L..39B}. To refine the LTE parameters, we optimized the initial CLASS/Weeds model by CLASS/ADJUST using the Powell algorithm \citep{Powell64} implemented in the SciPy optimization library. Because the uncertainties are dominated not only by the fitting procedure, but also by line blending, baseline effects, and the simplifying assumptions of the LTE model, we adopted a conservative uncertainty of 20\% for the derived $T_{\rm ex}$ and $N_{\rm t}$ values.

Although line intensities peak near the continuum maximum, a detailed line-by-line inspection revealed that most GA transitions in this region suffer from significant blending with neighboring species, making reliable identification and fitting particularly challenging. This severe blending is primarily due to the strongest Doppler broadening toward the center, where lines are most significantly widened and overlapping. To mitigate this issue, spectra were extracted from slightly offset positions (Fig.~\ref{fig:1} (b)), where most of the GA lines are less affected by blending and display clean Gaussian profiles with a high signal-to-noise ratios.

\begin{figure}[!htp]
    \centering
    \includegraphics[width=0.48\textwidth]{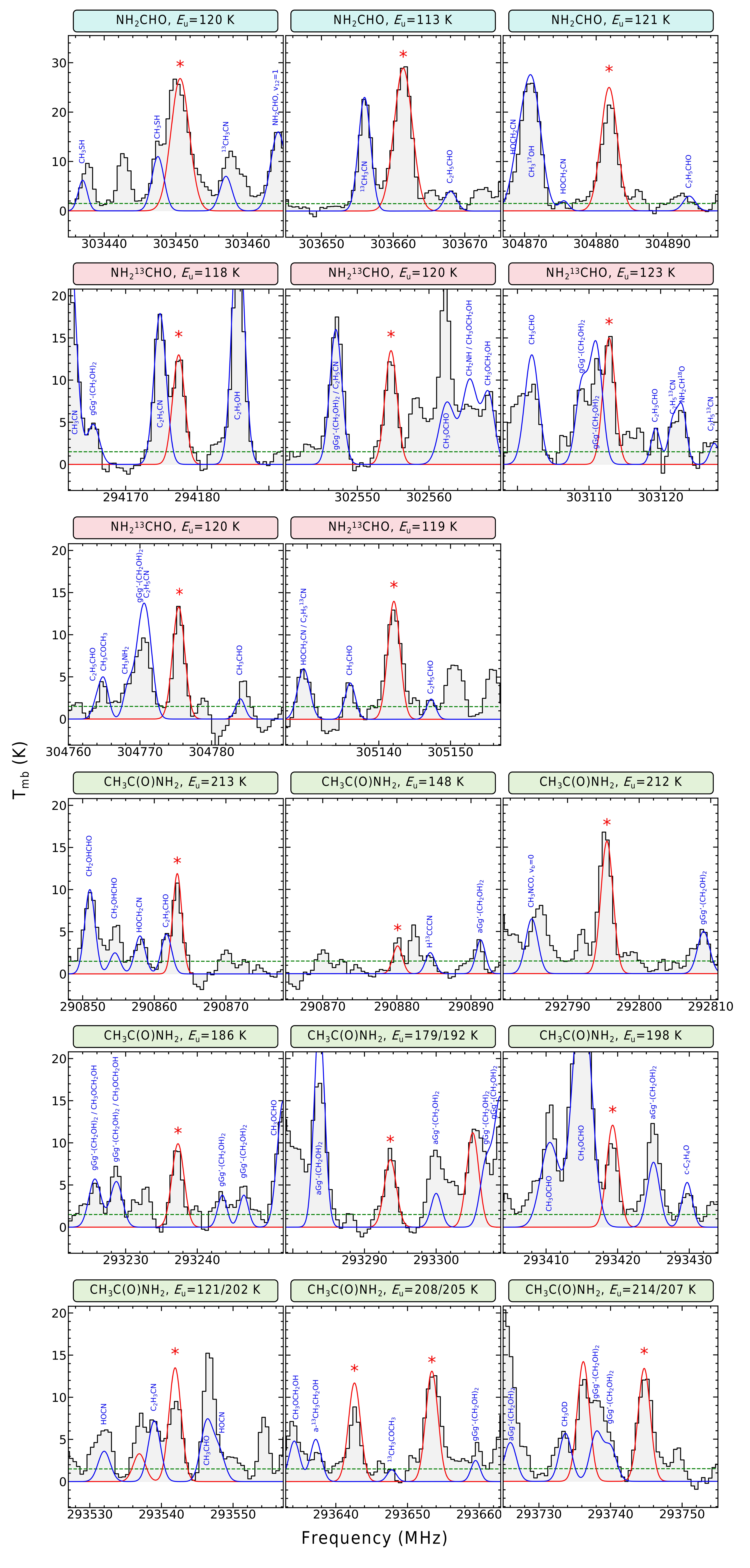}
    \caption{Observed (gray) and best-fit LTE-modeled (red) spectra of NH$_{2}$CHO, NH$_{2}$$^{13}$CHO, and CH$_{3}$C(O)NH$_{2}$ toward G358.93 MM1. All other plotting conventions are the same as in Fig.~\ref{fig:2}.}
    \label{fig:3}
\end{figure}

\begin{table}
\centering
\caption{Molecular parameters derived from LTE fits using CLASS, and corresponding abundances relative to H$_{2}$.}
\label{tab:1}
\begin{tabular}{cccc}
\hline
Species & $T_{\rm ex}$ & $N_{\rm t}$ & $\chi$$_{\rm H_{2}}$$^{\rm a}$ \\
&  (K) & (10$^{14}$ cm$^{-2}$) & ($10^{-10}$) \\
\hline
NH$_{2}$CHO$^{\rm b}$         & 124$\pm$25    & 18.8$\pm$3.8    & 2.4$\pm$0.7    \\
NH$_{2}$CHO$^{\rm c}$         & -             & 170.8$\pm$45.4   & 21.9$\pm$6.2   \\
NH$_{2}$$^{13}$CHO            & 112$\pm$22    & 6.1$\pm$1.2      & 0.8$\pm$0.2    \\
CH$_{3}$C(O)NH$_{2}$          & 171$\pm$34    & 88.1$\pm$17.6    & 11.3$\pm$3.1   \\
$syn$-HOCH$_{2}$C(O)NH$_{2}$  & 102$\pm$20    & 5.3$\pm$1.1      & 0.7$\pm$0.2    \\
\hline
\end{tabular}
        \tablefoot{\\
       $^{\rm a}$ Molecular abundances relative to H$_{2}$. The adopted $N$(H$_{2}$) and its derivation are described in Appendix ~\ref{app:E}.\\
       $^{\rm b}$ “Best-fit column density of NH$_{2}$CHO” refers to the parent isotopologue; optical-depth effects are discussed in the text.\\
       $^{\rm c}$ To mitigate the optical-depth effects, the NH$_{2}$CHO column density was derived from NH$_{2}$$^{13}$CHO, scaled by the ratio $^{12}$C/$^{13}$C = 28$\pm$5 \citep{2023A&A...670A..98Y}. 
       }
\end{table}

\subsection{Molecular line identifications} \label{sec:sec3.2}

Given the line-rich spectrum of G358.93 MM1, we constructed a full source model (Appendix ~\ref{app:B}) including the molecules reported in previous studies and the species identified in this work, in order to assess potential blends and overlaps in the candidate GA lines. The spectroscopic entries used in this model were retrieved from three main databases: CDMS \citep{Mul01, Mul05}, the Jet Propulsion Laboratory database \citep[JPL\footnote{\url{https://spec.jpl.nasa.gov}},] []{Pic98}, and the Lille Spectroscopic Database \citep[LSD\footnote{\url{https://lsd.univ-lille.fr}},] []{Mot25}. A complete list of all detected species is provided in Table B.1, together with the LTE parameters adopted for the full source model, including excitation temperature ($T_{\rm ex}$), column density ($N_{\rm t}$), line width (FWHM), and velocity offset ($V_{\rm off}$). For species other than GA, AA, and FA, these parameters were adjusted manually to best reproduce the observed spectrum. This full source model provides the basis for assessing possible blending and contamination in the candidate GA features.

Within this framework, GA is identified toward G358.93 MM1 through seven clean spectral features (Fig.~\ref{fig:2}), six of which are detected above the 3$\sigma$ level and three above 5$\sigma$ level. Some of these clean features consist of multiple overlapping GA transitions at high $K$. Taken together, the clean features comprise 17 transitions associated with 7 distinct $J$ values (see Table A.2 in Appendix A) with upper-level energies ($E_{\rm u}$) ranging from 108 to 389 K. All of these clean transitions are $R$-branch lines, including 7 $a$-type and 10 $b$-type transitions. Their relative intensities are well reproduced by the LTE model, no stronger clean GA lines expected within the covered bands are missing from the observed spectrum. However, given the limited bandwidth ($\sim$8 GHz), the modest number of securely detected transitions, and the high line density of this warm HMC, a coincidental assignment cannot be excluded with the confidence required for a firm detection claim. We therefore regard the identification as tentative. Under the LTE assumption, the best-fit model yields a column density of $N_{\rm t}$(GA) = (5.3$\pm$1.1)$\times$10$^{14} \rm cm^{-2}$ and an excitation temperature of $T_{\rm ex}$(GA) = 102$\pm$20 K, with the linewidth of GA set to 1.4 km s$^{-1}$.

As an independent consistency check on the best-fit LTE solution, we performed a rotational-diagram analysis based on seven unblended transitions of GA (Appendix ~\ref{app:C}). The derived rotational temperature, $T_{\rm rot}$ = 107$\pm$21 K, is in good agreement with the excitation temperature obtained from the best-fit, supporting the reliability of the adopted LTE treatment.

We also assessed the robustness of the continuum subtraction by comparing our CASA-based procedure with STATCONT \citep{Sanchez18}, an automated tool designed to determine the continuum level in line-rich spectra without manual intervention. The spectra produced by the two methods are nearly identical. The derived LTE parameters from the STATCONT-based spectra are also consistent with those from the CASA-based approach, although the column density derived from the STATCONT-based analysis is approximately 30\% lower than that from the CASA-based approach (see Appendix ~\ref{app:D} for details). This comparison indicates that the tentative GA identification is not driven by the particular continuum-subtraction method adopted.

Chemically related amides, FA and AA, were also examined. For FA, three unblended emission lines of the parent isotopologue and five of its $^{13}$C isotopologue (NH$_{2}$$^{13}$CHO) were detected (Fig.~\ref{fig:3}, Table A.2). In the analysis, the line widths were fixed to the average values of 2.8 km s$^{-1}$ for the FA lines and 2.0 km s$^{-1}$ for the NH$_{2}$$^{13}$CHO lines to derive the column densities and excitation temperatures. The derived parameters are $N_{\rm t}$(FA) = (18.8$\pm$3.8)$\times$$10^{14} \rm cm^{-2}$ with $T_{\rm ex}$(FA) = 124$\pm$25 K, and $N_{\rm t}$(NH$_{2}$$^{13}$CHO) = (6.1$\pm$1.2)$\times$$10^{14} \rm cm^{-2}$ with $T_{\rm ex}$(NH$_{2}$$^{13}$CHO) = 112$\pm$22 K. Because the FA lines are substantially optically thick, yielding an unrealistically low apparent $^{12}$C/$^{13}$C ratio of 3.1, the optically thin $^{13}$C isotopologue was used to infer the total FA column density $N_{\rm t}$(FA), adopting a scaling factor of $^{12}$C/$^{13}$C = 28$\pm$5 \citep{2023A&A...670A..98Y}. For AA, ten unblended lines with $E_{\rm u}$=148-213 K were identified, and their line widths were fixed to the average value of 2.1 km s$^{-1}$ in the analysis, giving $N_{\rm t}$(AA)= (88.1$\pm$17.6)$\times$10$^{14} \rm cm^{-2}$ and $T_{\rm ex}$(AA) = 171$\pm$34 K. All derived parameters are summarized in Table \ref{tab:1}. From these measurements, the molecular abundance ratios in G358.93 MM1 are FA/AA$\sim$1.9$\pm$0.6, AA/GA$\sim$16.6$\pm$4.8, and FA/GA$\sim$32.3$\pm$10.9. These ratios serve as key diagnostics for comparing amide chemistry across distinct interstellar environments and for testing chemical network models (see Section \ref{sec:sec4.1}).

\begin{figure*}[!htp]
    \centering
    \includegraphics[width=0.8\textwidth]{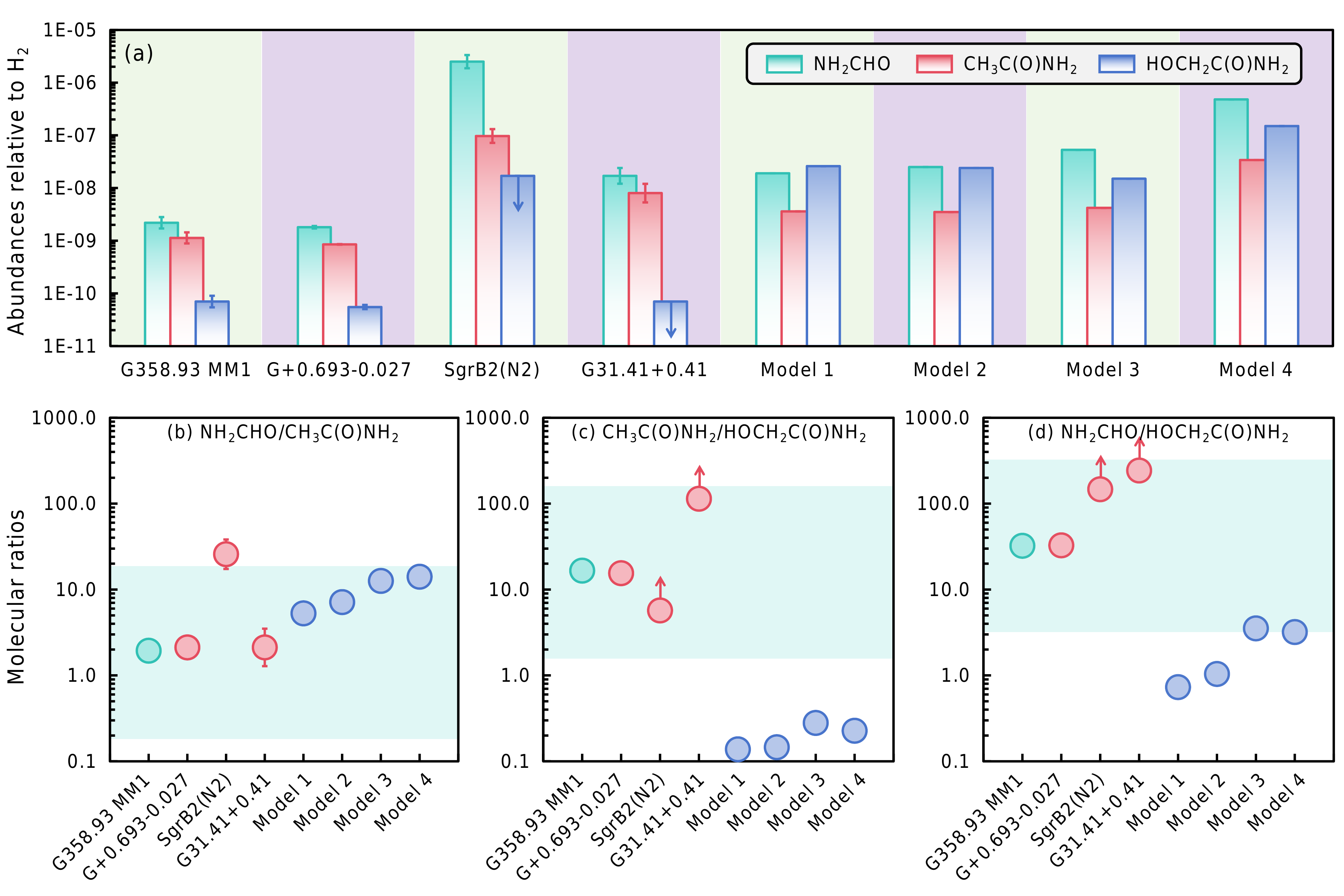}
    \caption{(a) Abundances of FA, AA, and GA relative to H$_{2}$ in G358.93 MM1, compared with other sources and model predictions. Model labels: (1-3) = Fast/Medium/Slow warm-up models from Garrod et al. \citep{2022ApJS..259....1G}; and (4) = the model of Sanz et al. \citep{2020ApJ...899...65S}. (b)-(d) Abundance ratios FA/AA, AA/GA, and FA/GA. Arrows denote lower limits. The cyan shaded regions indicate the uncertainty ranges (one order of magnitude) for the ratios derived in this work.}
    \label{fig:4}
\end{figure*}

\subsection{Spatial distribution} \label{sec:sec3.3}

The spatial distributions of FA, AA, and GA were analyzed using the Cube Analysis and Rendering Tool for Astronomy \citep[CARTA;][]{Com21}. Only transitions verified to be free from significant blending were utilized to generate the integrated-intensity (moment 0) maps. As shown in Fig.~\ref{fig:1}(c)-(e), the emissions of both FA and AA are nearly co-spatial with the 1 mm continuum peak of G358.93 MM1, consistent with previous findings for amide-bearing species \citep{2025A&A...702A..35D}. The newly obtained GA map is compact and coincident with both the FA and AA emission peaks as well as the continuum maximum. The close morphological correspondence among all three amides indicates that they likely share related formation environments or interconnected chemical pathways in the hot-core region.

\section{Discussion} \label{sec:sec4}
\subsection{Comparison with other sources and chemical models} \label{sec:sec4.1}

To place the amide chemistry of HMC G358.93 MM1 in context, the derived molecular abundances and abundance ratios have been compared with those measured in other sources and with predictions from chemical models. All abundances are normalized to H$_{2}$. The adopted $N$(H$_{2}$) and its derivation are described in Appendix ~\ref{app:E}. For reference, observational data were compiled for the molecular cloud G+0.693-0.027 \citep{2023ApJ...953L..20R, 2023MNRAS.523.1448Z} and two hot molecular cores including Sgr B2(N2) \citep{2020A&A...639A.135S, 2017A&A...601A..49B} and G31.41+0.41 \citep{2021A&A...653A.129C}. Model predictions were drawn from Sanz et al. and Garrod et al.\citep{2020ApJ...899...65S, 2022ApJS..259....1G}.

Across all sources, a consistent abundance hierarchy is apparent: AA and FA are substantially more abundant than GA, while FA is typically comparable to or slightly higher than AA \citep{2023ApJ...953L..20R, 2020A&A...639A.135S, 2021A&A...653A.129C, 2023MNRAS.523.1448Z, 2017A&A...601A..49B}. G358.93 MM1 follows this overall ordering, although its FA/AA ratio lies toward the lower end of the hot-core distribution. In contrast, significantly higher values—such as FA/AA $\approx$ 25.8 in Sgr B2(N2) \citep{2017A&A...601A..49B}—suggest that source-to-source variations may reflect a combination of optical-depth effects, excitation temperature differences, beam dilution, and time-dependent chemistry along distinct warm-up tracks \citep{2022ApJS..259....1G}. Despite these differences, the abundance ratios among FA, AA, and GA generally agree within an order of magnitude, suggesting that the underlying amide chemistry is broadly consistent across environments.

The most striking discrepancy arises for GA, whose observed abundance is systematically lower than predicted by current astrochemical models. While models \citep{2020ApJ...899...65S, 2022ApJS..259....1G} tend to produce GA at levels comparable to or exceeding AA, observations of both G358.93 MM1 and G+0.693-0.027 show it to be at least one order of magnitude lower than AA \citep[this work and][]{2023ApJ...953L..20R}. This tension likely reflects an overestimation of GA formation efficiency and/or an underrepresentation of its destruction processes in current networks—potentially due to optimistic diffusion barriers, incomplete treatment of OH/H-abstraction reactions, or neglected photolytic pathways \citep{2020ApJ...899...65S, 2022ApJS..259....1G}.

\subsection{Chemical linkages of GA with FA and AA} \label{sec:sec4.2}

Since the first detection of GA in G+0.693-0.027 \citep{2023ApJ...953L..20R}, its chemical relationship to AA has been a central topic of discussion. Structurally, GA is the $\alpha$-hydroxy derivative of AA (–CH$_{3}$ $\rightarrow$ –CH$_{2}$OH), naturally suggesting a shared formation network. GA can form via the OH + NH$_{2}$COCH$_{2}$ reaction, where the NH$_{2}$COCH$_{2}$ radical has been experimentally demonstrated to arise from H-abstraction at the methyl site of AA \citep{2020PCCP...22.6192H}. An alternative formation route proposed by Garrod et al. \citep{2008ApJ...682..283G} begins with H-abstraction from FA to produce the NH$_{2}$CO radical, which then reacts with CH$_{2}$OH to yield GA. These pathways collectively link GA formation to both FA and AA through common intermediate radicals.

Observationally, our ALMA data toward G358.93 MM1 are consistent with this precursor-based framework. Both AA and FA are substantially more abundant than GA (AA/GA$\sim$16.6, FA/GA$\sim$32.3; Fig.~\ref{fig:4}), providing ample reservoirs of precursor radicals under warm-up conditions. Moreover, the close spatial correspondence among GA, AA, and FA (Fig.~\ref{fig:1}) further supports a shared chemical environment. The similarity of the FA/GA and AA/GA ratios between G358 MM1 and G+0.693-0.027 \citep{2023ApJ...953L..20R, 2023MNRAS.523.1448Z} suggests that analogous amide-forming networks may operate across both quiescent and hot-core regimes.

Nevertheless, with only two confirmed GA detections to date, any conclusions regarding its precise formation mechanisms remain preliminary. Expanded surveys targeting sources with different warm-up timescales, UV and cosmic-ray irradiation levels, and ice compositions—together with models that incorporate refined diffusion barriers, abstraction/destruction kinetics, and reactive-desorption efficiencies—will be crucial to constrain the chemical linkage of GA to FA and AA.

\section{Conclusions} \label{sec:sec5}
Using ALMA 1 mm data, we present the tentative detection of $syn$-glycolamide ($syn$-HOCH$_{2}$C(O)NH$_{2}$, GA) toward a hot molecular core, G358.93–0.03 MM1, through seven unblended transitions with upper-level energies ranging from 108 to 389 K. Confirmation of this tentative identification will require future observations with broader frequency coverage to reduce the likelihood of coincidental line blending in this line-rich source. The derived abundance of GA relative to H$_{2}$ is (6.8$\pm$2.0)$\times$10$^{-11}$. In the same source, acetamide (CH$_{3}$C(O)NH$_{2}$, AA) and formamide (NH$_{2}$CHO, FA) were also detected. The abundance ratios FA/GA and AA/GA in G358.93-0.03 MM1 closely resemble those measured in the molecular cloud G+0.693-0.027. This similarities suggest that related amide-forming networks operate in different environments, though the small number of GA detections precludes firm conclusions. Notably, current astrochemical models appear to overproduce GA, implying that its formation may be overestimated and/or key destruction pathways are missing.

This detection extends the known occurrence of GA from Galactic Center clouds to warm, dense protostellar environments, demonstrating that complex amides—and by extension, amino acid precursors—can survive under the physical conditions characteristic of massive star formation. Continued laboratory, observational, and modeling efforts will be essential to constrain the formation routes and assess the broader astrochemical role of such species.

\section*{Data availability} \label{sec:sec6}
The ALMA data of G358.93-0.03 is used in this paper. The raw dataset and documentation can be downloaded from \url{https://almascience.nrao.edu/aq/}. The derived data underlying this article are available in the article and in its online supplementary material on \href{https://doi.org/10.5281/zenodo.19394238}{Zenodo}.

\begin{acknowledgements}

This work makes use of the following ALMA data: ADS/JAO.ALMA\#2019.1.00768.S. ALMA is a partnership of ESO (representing its member states), NSF (USA), and NINS (Japan), together with NRC (Canada), MOST and ASIAA (Taiwan, China), and KASI (Republic of Korea), in cooperation with the Republic of Chile. The Joint ALMA Observatory is operated by ESO, AUI/NRAO, and NAOJ. We are grateful for support from the National SKA Program of China (No. 2025SKA0120100), National Natural Science Foundation of China (Grant No. W2512014), Fundamental Research Funds for the Central Universities (Grant No. 2025CDJ-IAISYB-060), and Postdoctoral Fellows Excellence Support Program (Grant No. 2404013554893087).
      
\end{acknowledgements}

\bibliographystyle{aa} 
\bibliography{GA} 

\begin{thebibliography}{74}
\expandafter\ifx\csname natexlab\endcsname\relax\def\natexlab#1{#1}\fi

\bibitem[{{Altwegg} {et~al.}(2016){Altwegg}, {Balsiger}, {Bar-Nun},
  {Berthelier}, {Bieler}, {Bochsler}, {Briois}, {Calmonte}, {Combi}, {Cottin},
  {De Keyser}, {Dhooghe}, {Fiethe}, {Fuselier}, {Gasc}, {Gombosi}, {Hansen},
  {Haessig}, {Ja ckel}, {Kopp}, {Korth}, {Le Roy}, {Mall}, {Marty}, {Mousis},
  {Owen}, {Reme}, {Rubin}, {Semon}, {Tzou}, {Waite}, \&
  {Wurz}}]{2016SciA....2E0285A}
{Altwegg}, K., {Balsiger}, H., {Bar-Nun}, A., {et~al.} 2016, Science Advances,
  2, e1600285

\bibitem[{Bakri {et~al.}(2002)Bakri, Demaison, Kleiner, Margulès, Møllendal,
  Petitprez, \& Wlodarczak}]{BAKRI2002312}
Bakri, B., Demaison, J., Kleiner, I., {et~al.} 2002, Journal of Molecular
  Spectroscopy, 215, 312

\bibitem[{{Belloche} {et~al.}(2017){Belloche}, {Meshcheryakov}, {Garrod},
  {Ilyushin}, {Alekseev}, {Motiyenko}, {Margul{\`e}s}, {M{\"u}ller}, \&
  {Menten}}]{2017A&A...601A..49B}
{Belloche}, A., {Meshcheryakov}, A.~A., {Garrod}, R.~T., {et~al.} 2017, \aap,
  601, A49

\bibitem[{Bergantini {et~al.}(2025)Bergantini, Wang, Antonov, Batrakova,
  Tuchin, \& Kaiser}]{Bergantini25}
Bergantini, A., Wang, J., Antonov, I., {et~al.} 2025, ACS Central Science

\bibitem[{{Blanco} {et~al.}(2006){Blanco}, {L{\'o}pez}, {Lesarri}, \&
  {Alonso}}]{2006JAChS.12812111B}
{Blanco}, S., {L{\'o}pez}, J.~C., {Lesarri}, A., \& {Alonso}, J.~L. 2006,
  Journal of the American Chemical Society, 128, 12111

\bibitem[{{Bonfand} {et~al.}(2019){Bonfand}, {Belloche}, {Garrod}, {Menten},
  {Willis}, {St{\'e}phan}, \& {M{\"u}ller}}]{Bon19}
{Bonfand}, M., {Belloche}, A., {Garrod}, R.~T., {et~al.} 2019, \aap, 628, A27

\bibitem[{{Brogan} {et~al.}(2019){Brogan}, {Hunter}, {Towner}, {McGuire},
  {MacLeod}, {Gurwell}, {Cyganowski}, {Brand}, {Burns}, {Caratti o Garatti},
  {Chen}, {Chibueze}, {Hirano}, {Hirota}, {Kim}, {Kramer}, {Linz}, {Menten},
  {Remijan}, {Sanna}, {Sobolev}, {Sridharan}, {Stecklum}, {Sugiyama}, {Surcis},
  {Van der Walt}, {Volvach}, \& {Volvach}}]{2019ApJ...881L..39B}
{Brogan}, C.~L., {Hunter}, T.~R., {Towner}, A.~P.~M., {et~al.} 2019, \apjl,
  881, L39

\bibitem[{{Burton} {et~al.}(2012){Burton}, {Dworkin}, {Callahan}, {Glavin}, \&
  {Elsila}}]{2012cosp...39..264B}
{Burton}, A., {Dworkin}, J., {Callahan}, M., {Glavin}, D., \& {Elsila}, J.
  2012, in 39th COSPAR Scientific Assembly, Vol.~39, 264

\bibitem[{{Ceccarelli} {et~al.}(2000){Ceccarelli}, {Loinard}, {Castets},
  {Faure}, \& {Lefloch}}]{2000A&A...362.1122C}
{Ceccarelli}, C., {Loinard}, L., {Castets}, A., {Faure}, A., \& {Lefloch}, B.
  2000, \aap, 362, 1122

\bibitem[{{Chen} {et~al.}(2020){Chen}, {Sobolev}, {Ren}, {Parfenov}, {Breen},
  {Ellingsen}, {Shen}, {Li}, {MacLeod}, {Baan}, {Brogan}, {Hirota}, {Hunter},
  {Linz}, {Menten}, {Sugiyama}, {Stecklum}, {Gong}, \&
  {Zheng}}]{2020NatAs...4.1170C}
{Chen}, X., {Sobolev}, A.~M., {Ren}, Z.-Y., {et~al.} 2020, Nature Astronomy, 4,
  1170

\bibitem[{{Colzi} {et~al.}(2021){Colzi}, {Rivilla}, {Beltr{\'a}n},
  {Jim{\'e}nez-Serra}, {Mininni}, {Melosso}, {Cesaroni}, {Fontani},
  {Lorenzani}, {S{\'a}nchez-Monge}, {Viti}, {Schilke}, {Testi}, {Alonso}, \&
  {Kolesnikov{\'a}}}]{2021A&A...653A.129C}
{Colzi}, L., {Rivilla}, V.~M., {Beltr{\'a}n}, M.~T., {et~al.} 2021, \aap, 653,
  A129

\bibitem[{{Combes} {et~al.}(1996){Combes}, {Q-Rieu}, \&
  {Wlodarczak}}]{1996A&A...308..618C}
{Combes}, F., {Q-Rieu}, N., \& {Wlodarczak}, G. 1996, \aap, 308, 618

\bibitem[{{Comrie} {et~al.}(2021){Comrie}, {Wang}, {Hsu}, {Moraghan}, {Harris},
  {Pang}, {Pi{\'n}ska}, {Chiang}, {Chang}, {Hwang}, {Jan}, {Lin}, \&
  {Simmonds}}]{Com21}
{Comrie}, A., {Wang}, K.-S., {Hsu}, S.-C., {et~al.} 2021, {CARTA: The Cube
  Analysis and Rendering Tool for Astronomy}

\bibitem[{{Cronin} \& {Pizzarello}(1983)}]{1983AdSpR...3i...5C}
{Cronin}, J.~R. \& {Pizzarello}, S. 1983, Advances in Space Research, 3, 5

\bibitem[{{Cunningham} {et~al.}(2007){Cunningham}, {Jones}, {Godfrey}, {Cragg},
  {Bains}, {Burton}, {Calisse}, {Crighton}, {Curran}, {Davis}, {Dempsey},
  {Fulton}, {Hidas}, {Hill}, {Kedziora-Chudczer}, {Minier}, {Pracy}, {Purcell},
  {Shobbrook}, \& {Travouillon}}]{2007MNRAS.376.1201C}
{Cunningham}, M.~R., {Jones}, P.~A., {Godfrey}, P.~D., {et~al.} 2007, \mnras,
  376, 1201

\bibitem[{{Demyk} {et~al.}(2004){Demyk}, {Wlodarczak}, \&
  {Dartois}}]{2004sf2a.conf..493D}
{Demyk}, K., {Wlodarczak}, G., \& {Dartois}, E. 2004, in SF2A-2004: Semaine de
  l'Astrophysique Francaise, ed. F.~{Combes}, D.~{Barret}, T.~{Contini},
  F.~{Meynadier}, \& L.~{Pagani}, 493

\bibitem[{{Duan} {et~al.}(2025{\natexlab{a}}){Duan}, {Gou}, {Liu}, {Xu}, {Xu},
  {Lan}, {Wang}, {Pagani}, {Quan}, {Wang}, {Liu}, \&
  {He}}]{2025ApJ...988...95D}
{Duan}, C., {Gou}, Q., {Liu}, T., {et~al.} 2025{\natexlab{a}}, \apj, 988, 95

\bibitem[{{Duan} {et~al.}(2025{\natexlab{b}}){Duan}, {Xu}, {Gou}, {Liu},
  {Pagani}, {Xu}, {Wang}, {Liu}, {Kang}, {He}, \& {Jiao}}]{2025A&A...702A..35D}
{Duan}, C., {Xu}, X., {Gou}, Q., {et~al.} 2025{\natexlab{b}}, \aap, 702, A35

\bibitem[{{Gardner} {et~al.}(1980){Gardner}, {Godfrey}, \&
  {Williams}}]{1980MNRAS.193..713G}
{Gardner}, F.~F., {Godfrey}, P.~D., \& {Williams}, D.~R. 1980, \mnras, 193, 713

\bibitem[{{Garrod} {et~al.}(2022){Garrod}, {Jin}, {Matis}, {Jones}, {Willis},
  \& {Herbst}}]{2022ApJS..259....1G}
{Garrod}, R.~T., {Jin}, M., {Matis}, K.~A., {et~al.} 2022, \apjs, 259, 1

\bibitem[{{Garrod} {et~al.}(2008){Garrod}, {Widicus Weaver}, \&
  {Herbst}}]{2008ApJ...682..283G}
{Garrod}, R.~T., {Widicus Weaver}, S.~L., \& {Herbst}, E. 2008, \apj, 682, 283

\bibitem[{Goldsmith \& Langer(1999)}]{goldsmith1999}
Goldsmith, P.~F. \& Langer, W.~D. 1999, The Astrophysical Journal, 517, 209

\bibitem[{{Haupa} {et~al.}(2020){Haupa}, {Ong}, \& {Lee}}]{2020PCCP...22.6192H}
{Haupa}, K.~A., {Ong}, W.-S., \& {Lee}, Y.-P. 2020, Physical Chemistry Chemical
  Physics (Incorporating Faraday Transactions), 22, 6192

\bibitem[{{Heineking} \& {Dreizler}(1993)}]{1993ZNatA..48..787H}
{Heineking}, N. \& {Dreizler}, H. 1993, Zeitschrift Naturforschung Teil A, 48,
  787

\bibitem[{{Hirota} {et~al.}(1974){Hirota}, {Sugisaki}, {Nielsen}, \&
  {S{\o}rensen}}]{1974JMoSp..49..251H}
{Hirota}, E., {Sugisaki}, R., {Nielsen}, C.~J., \& {S{\o}rensen}, G.~O. 1974,
  Journal of Molecular Spectroscopy, 49, 251

\bibitem[{Ilyushin {et~al.}(2006)Ilyushin, Alekseev, Demaison, \&
  Kleiner}]{ILYUSHIN2006127}
Ilyushin, V., Alekseev, E., Demaison, J., \& Kleiner, I. 2006, Journal of
  Molecular Spectroscopy, 240, 127

\bibitem[{{Ilyushin} {et~al.}(2004){Ilyushin}, {Alekseev}, {Dyubko}, {Kleiner},
  \& {Hougen}}]{2004JMoSp.227..115I}
{Ilyushin}, V.~V., {Alekseev}, E.~A., {Dyubko}, S.~F., {Kleiner}, I., \&
  {Hougen}, J.~T. 2004, Journal of Molecular Spectroscopy, 227, 115

\bibitem[{{Jim{\'e}nez-Serra} {et~al.}(2020){Jim{\'e}nez-Serra},
  {Mart{\'\i}n-Pintado}, {Rivilla}, {Rodr{\'\i}guez-Almeida}, {Alonso Alonso},
  {Zeng}, {Cocinero}, {Mart{\'\i}n}, {Requena-Torres}, {Mart{\'\i}n-Domenech},
  \& {Testi}}]{2020AsBio..20.1048J}
{Jim{\'e}nez-Serra}, I., {Mart{\'\i}n-Pintado}, J., {Rivilla}, V.~M., {et~al.}
  2020, Astrobiology, 20, 1048

\bibitem[{{Jim{\'e}nez-Serra} {et~al.}(2016){Jim{\'e}nez-Serra}, {Vasyunin},
  {Caselli}, {Marcelino}, {Billot}, {Viti}, {Testi}, {Vastel}, {Lefloch}, \&
  {Bachiller}}]{2016ApJ...830L...6J}
{Jim{\'e}nez-Serra}, I., {Vasyunin}, A.~I., {Caselli}, P., {et~al.} 2016,
  \apjl, 830, L6

\bibitem[{{Jones} {et~al.}(2007){Jones}, {Cunningham}, {Godfrey}, \&
  {Cragg}}]{2007MNRAS.374..579J}
{Jones}, P.~A., {Cunningham}, M.~R., {Godfrey}, P.~D., \& {Cragg}, D.~M. 2007,
  \mnras, 374, 579

\bibitem[{{Kauffmann} {et~al.}(2008){Kauffmann}, {Bertoldi}, {Bourke}, {Evans},
  \& {Lee}}]{Kau08}
{Kauffmann}, J., {Bertoldi}, F., {Bourke}, T.~L., {Evans}, N.~J., I., \& {Lee},
  C.~W. 2008, \aap, 487, 993

\bibitem[{{Kojima} {et~al.}(1987){Kojima}, {Yano}, {Nakagawa}, \&
  {Tsunekawa}}]{1987JMoSp.122..408K}
{Kojima}, T., {Yano}, E., {Nakagawa}, K., \& {Tsunekawa}, S. 1987, Journal of
  Molecular Spectroscopy, 122, 408

\bibitem[{{Kryvda} {et~al.}(2009){Kryvda}, {Gerasimov}, {Dyubko}, {Alekseev},
  \& {Motiyenko}}]{2009JMoSp.254...28K}
{Kryvda}, A.~V., {Gerasimov}, V.~G., {Dyubko}, S.~F., {Alekseev}, E.~A., \&
  {Motiyenko}, R.~A. 2009, Journal of Molecular Spectroscopy, 254, 28

\bibitem[{{Kuan} {et~al.}(2003){Kuan}, {Charnley}, {Huang}, {Tseng}, \&
  {Kisiel}}]{2003ApJ...593..848K}
{Kuan}, Y.-J., {Charnley}, S.~B., {Huang}, H.-C., {Tseng}, W.-L., \& {Kisiel},
  Z. 2003, \apj, 593, 848

\bibitem[{{Kukolich} \& {Nelson}(1971)}]{1971CPL....11..383K}
{Kukolich}, S.~G. \& {Nelson}, A.~C. 1971, Chemical Physics Letters, 11, 383

\bibitem[{Liu {et~al.}(2002)Liu, Girart, Remijan, \& Snyder}]{liu2002formic}
Liu, S.-Y., Girart, J., Remijan, A., \& Snyder, L. 2002, The Astrophysical
  Journal, 576, 255

\bibitem[{{Manna} \& {Pal}(2023)}]{2023Ap&SS.368...33M}
{Manna}, A. \& {Pal}, S. 2023, \apss, 368, 33

\bibitem[{{Manna} \& {Pal}(2024)}]{2024RAA....24g5014M}
{Manna}, A. \& {Pal}, S. 2024, Research in Astronomy and Astrophysics, 24,
  075014

\bibitem[{{Manna} {et~al.}(2024){Manna}, {Pal}, \&
  {Viti}}]{2024MNRAS.533.1143M}
{Manna}, A., {Pal}, S., \& {Viti}, S. 2024, \mnras, 533, 1143

\bibitem[{{Manna} {et~al.}(2023){Manna}, {Pal}, {Viti}, \&
  {Sinha}}]{2023MNRAS.525.2229M}
{Manna}, A., {Pal}, S., {Viti}, S., \& {Sinha}, S. 2023, \mnras, 525, 2229

\bibitem[{{Maret} {et~al.}(2011){Maret}, {Hily-Blant}, {Pety}, {Bardeau}, \&
  {Reynier}}]{Mar11}
{Maret}, S., {Hily-Blant}, P., {Pety}, J., {Bardeau}, S., \& {Reynier}, E.
  2011, \aap, 526, A47

\bibitem[{Maris(2004)}]{Maris04}
Maris, A. 2004, Phys. Chem. Chem. Phys., 6, 2611

\bibitem[{Marstokk \& Mollendal(1999)}]{Mar1999}
Marstokk, K. \& Mollendal, H. 1999, Acta Chemica Scandinavica, 53, 79

\bibitem[{{McMullin} {et~al.}(2007){McMullin}, {Waters}, {Schiebel}, {Young},
  \& {Golap}}]{Mcm07}
{McMullin}, J.~P., {Waters}, B., {Schiebel}, D., {Young}, W., \& {Golap}, K.
  2007, in Astronomical Society of the Pacific Conference Series, Vol. 376,
  Astronomical Data Analysis Software and Systems XVI, ed. R.~A. {Shaw},
  F.~{Hill}, \& D.~J. {Bell}, 127

\bibitem[{{Millar}(2020)}]{2020ChJCP..33..668M}
{Millar}, T.~J. 2020, Chinese Journal of Chemical Physics, 33, 668

\bibitem[{{Moskienko} \& {Diubko}(1991)}]{1991RaF....34..213M}
{Moskienko}, E.~M. \& {Diubko}, S.~F. 1991, Radiofizika, 34, 213

\bibitem[{{Motiyenko} \& {Margul{\`e}s}(2025)}]{Mot25}
{Motiyenko}, R.~A. \& {Margul{\`e}s}, L. 2025, \aap, 699, A348

\bibitem[{{Motiyenko} {et~al.}(2012){Motiyenko}, {Tercero}, {Cernicharo}, \&
  {Margul{\`e}s}}]{2012AA...548A..71M}
{Motiyenko}, R.~A., {Tercero}, B., {Cernicharo}, J., \& {Margul{\`e}s}, L.
  2012, \aap, 548, A71

\bibitem[{{M{\"u}ller} {et~al.}(2005){M{\"u}ller}, {Schl{\"o}der}, {Stutzki},
  \& {Winnewisser}}]{Mul05}
{M{\"u}ller}, H. S.~P., {Schl{\"o}der}, F., {Stutzki}, J., \& {Winnewisser}, G.
  2005, Journal of Molecular Structure, 742, 215

\bibitem[{{M{\"u}ller} {et~al.}(2001){M{\"u}ller}, {Thorwirth}, {Roth}, \&
  {Winnewisser}}]{Mul01}
{M{\"u}ller}, H.~S.~P., {Thorwirth}, S., {Roth}, D.~A., \& {Winnewisser}, G.
  2001, \aap, 370, L49

\bibitem[{{Pearce} {et~al.}(2018){Pearce}, {Tupper}, {Pudritz}, \&
  {Higgs}}]{2018AsBio..18..343P}
{Pearce}, B. K.~D., {Tupper}, A.~S., {Pudritz}, R.~E., \& {Higgs}, P.~G. 2018,
  Astrobiology, 18, 343

\bibitem[{{Perrero} {et~al.}(2025){Perrero}, {Alessandrini}, {Ye}, {Puzzarini},
  \& {Rimola}}]{Perrero25}
{Perrero}, J., {Alessandrini}, S., {Ye}, H., {Puzzarini}, C., \& {Rimola}, A.
  2025, \aap, 698, A51

\bibitem[{{Pickett} {et~al.}(1998){Pickett}, {Poynter}, {Cohen}, {Delitsky},
  {Pearson}, \& {M{\"u}ller}}]{Pic98}
{Pickett}, H.~M., {Poynter}, R.~L., {Cohen}, E.~A., {et~al.} 1998, \jqsrt, 60,
  883

\bibitem[{{Pizzarello} \& {Weber}(2004)}]{2004Sci...303.1151P}
{Pizzarello}, S. \& {Weber}, A.~L. 2004, Science, 303, 1151

\bibitem[{{Potiszil} {et~al.}(2023){Potiszil}, {Ota}, {Yamanaka}, {Sakaguchi},
  {Kobayashi}, {Tanaka}, {Kunihiro}, {Kitagawa}, {Abe}, {Miyazaki}, {Nakato},
  {Nakazawa}, {Nishimura}, {Okada}, {Saiki}, {Tanaka}, {Terui}, {Tsuda},
  {Usui}, {Watanabe}, {Yada}, {Yogata}, {Yoshikawa}, \&
  {Nakamura}}]{2023NatCo..14.1482P}
{Potiszil}, C., {Ota}, T., {Yamanaka}, M., {et~al.} 2023, Nature
  Communications, 14, 1482

\bibitem[{Powell(1964)}]{Powell64}
Powell, M. J.~D. 1964, The Computer Journal, 7, 155

\bibitem[{Qin {et~al.}(2010)Qin, Wu, Huang, Zhao, Li, Wang, \&
  Chen}]{qin2010high}
Qin, S.-L., Wu, Y., Huang, M., {et~al.} 2010, The Astrophysical Journal, 711,
  399

\bibitem[{Remijan {et~al.}(2003)Remijan, Snyder, Friedel, Liu, \&
  Shah}]{remijan2003}
Remijan, A., Snyder, L.~E., Friedel, D.~N., Liu, S.-Y., \& Shah, R.~Y. 2003,
  The Astrophysical Journal, 590, 314

\bibitem[{{Rivilla} {et~al.}(2023){Rivilla}, {Sanz-Novo}, {Jim{\'e}nez-Serra},
  {Mart{\'\i}n-Pintado}, {Colzi}, {Zeng}, {Meg{\'\i}as}, {L{\'o}pez-Gallifa},
  {Mart{\'\i}nez-Henares}, {Massalkhi}, {Tercero}, {de Vicente}, {Mart{\'\i}n},
  {San Andr{\'e}s}, {Requena-Torres}, \& {Alonso}}]{2023ApJ...953L..20R}
{Rivilla}, V.~M., {Sanz-Novo}, M., {Jim{\'e}nez-Serra}, I., {et~al.} 2023,
  \apjl, 953, L20

\bibitem[{Ruiz-Mirazo {et~al.}(2014)Ruiz-Mirazo, Briones, \& de~la
  Escosura}]{ruiz2014prebiotic}
Ruiz-Mirazo, K., Briones, C., \& de~la Escosura, A. 2014, Chemical reviews,
  114, 285

\bibitem[{{Sahu} {et~al.}(2020){Sahu}, {Liu}, {Das}, {Garai}, \&
  {Wakelam}}]{2020ApJ...899...65S}
{Sahu}, D., {Liu}, S.-Y., {Das}, A., {Garai}, P., \& {Wakelam}, V. 2020, \apj,
  899, 65

\bibitem[{{S{\'a}nchez-Monge} {et~al.}(2018){S{\'a}nchez-Monge}, {Schilke},
  {Ginsburg}, {Cesaroni}, \& {Schmiedeke}}]{Sanchez18}
{S{\'a}nchez-Monge}, {\'A}., {Schilke}, P., {Ginsburg}, A., {Cesaroni}, R., \&
  {Schmiedeke}, A. 2018, \aap, 609, A101

\bibitem[{{Sanz-Novo} {et~al.}(2022){Sanz-Novo}, {Alonso}, {Rivilla},
  {McGuire}, {Le{\'o}n}, {Mata}, {Jimenez-Serra}, \&
  {Mart{\'\i}n-Pintado}}]{2022A&A...666A.134S}
{Sanz-Novo}, M., {Alonso}, J.~L., {Rivilla}, V.~M., {et~al.} 2022, \aap, 666,
  A134

\bibitem[{{Sanz-Novo} {et~al.}(2020){Sanz-Novo}, {Belloche}, {Alonso},
  {Kolesnikov{\'a}}, {Garrod}, {Mata}, {M{\"u}ller}, {Menten}, \&
  {Gong}}]{2020A&A...639A.135S}
{Sanz-Novo}, M., {Belloche}, A., {Alonso}, J.~L., {et~al.} 2020, \aap, 639,
  A135

\bibitem[{{Shimonishi} {et~al.}(2021){Shimonishi}, {Izumi}, {Furuya}, \&
  {Yasui}}]{2021ApJ...922..206S}
{Shimonishi}, T., {Izumi}, N., {Furuya}, K., \& {Yasui}, C. 2021, \apj, 922,
  206

\bibitem[{{Snyder} {et~al.}(2005){Snyder}, {Lovas}, {Hollis}, {Friedel},
  {Jewell}, {Remijan}, {Ilyushin}, {Alekseev}, \&
  {Dyubko}}]{2005ApJ...619..914S}
{Snyder}, L.~E., {Lovas}, F.~J., {Hollis}, J.~M., {et~al.} 2005, \apj, 619, 914

\bibitem[{{Suenram} {et~al.}(2001){Suenram}, {Golubiatnikov}, {Leonov},
  {Hougen}, {Ortigoso}, {Kleiner}, \& {Fraser}}]{2001JMoSp.208..188S}
{Suenram}, R.~D., {Golubiatnikov}, G.~Y., {Leonov}, I.~I., {et~al.} 2001,
  Journal of Molecular Spectroscopy, 208, 188

\bibitem[{{Sz{\H{o}}ri} {et~al.}(2011){Sz{\H{o}}ri}, {J{\'o}j{\'a}rt},
  {Izs{\'a}k}, {Sz{\H{o}}ri}, {Csizmadia}, \& {Viskolcz}}]{2011PCCP...13.7449S}
{Sz{\H{o}}ri}, M., {J{\'o}j{\'a}rt}, B., {Izs{\'a}k}, R., {et~al.} 2011,
  Physical Chemistry Chemical Physics (Incorporating Faraday Transactions), 13,
  7449

\bibitem[{{Vorob'eva} \& {Dyubko}(1994)}]{1994RQE...37..155V}
{Vorob'eva}, E.~M. \& {Dyubko}, S.~F. 1994, Radiophysics and Quantum
  Electronics, 37, 155

\bibitem[{{Weber} \& {Pizzarello}(2006)}]{2006PNAS..10312713W}
{Weber}, A.~L. \& {Pizzarello}, S. 2006, Proceedings of the National Academy of
  Science, 103, 12713

\bibitem[{Wu(2009)}]{wu2009amino}
Wu, G. 2009, Amino acids, 37, 1

\bibitem[{{Yamaguchi} {et~al.}(2002){Yamaguchi}, {Hagiwara}, {Odashima},
  {Takagi}, \& {Tsunekawa}}]{2002JMoSp.215..144Y}
{Yamaguchi}, A., {Hagiwara}, S., {Odashima}, H., {Takagi}, K., \& {Tsunekawa},
  S. 2002, Journal of Molecular Spectroscopy, 215, 144

\bibitem[{{Yan} {et~al.}(2023){Yan}, {Henkel}, {Kobayashi}, {Menten}, {Gong},
  {Zhang}, {Yu}, {Yang}, {Xie}, \& {Wang}}]{2023A&A...670A..98Y}
{Yan}, Y.~T., {Henkel}, C., {Kobayashi}, C., {et~al.} 2023, \aap, 670, A98

\bibitem[{{Zeng} {et~al.}(2023){Zeng}, {Rivilla}, {Jim{\'e}nez-Serra}, {Colzi},
  {Mart{\'\i}n-Pintado}, {Tercero}, {de Vicente}, {Mart{\'\i}n}, \&
  {Requena-Torres}}]{2023MNRAS.523.1448Z}
{Zeng}, S., {Rivilla}, V.~M., {Jim{\'e}nez-Serra}, I., {et~al.} 2023, \mnras,
  523, 1448

\end{thebibliography}

\begin{appendix}

\onecolumn

\section{Spectroscopic Line Lists}
\label{app:A}

Spectroscopic line lists used in this work were retrieved from the CDMS database. Table \ref{tab:A1} summarizes the catalogs entries, molecular identifiers, and the primary laboratory or theoretical studies on which each entry is based. Table A.2, only available on \href{https://doi.org/10.5281/zenodo.19394238}{Zenodo}, lists spectroscopic parameters of all unblended or minimally blended transitions analyzed in this work, including rest frequencies, upper-state energies ($E_{\rm up}$), Einstein A-coefficients ($A_{\rm ij}$), degeneracies ($g_{\rm up}$), line width ($\Delta V$), and integrated intensity ($\int{T_{\rm MB}}dv$).

\begin{table}[!htp]
\centering
\caption{Summary of spectroscopic data sources and references used in this work.}
\label{tab:A1}
\begin{tabular}{l c c c}
\hline
\hline
Sources & Species & Catalog ID & Ref. \\
\hline
    \multirow{19}{*}{CDMS} & \multirow{7}{*}{Formamide} & \multirow{8}{*}{045512} & \citet{1971CPL....11..383K} \\
                                   & \multirow{7}{*}{$\rm NH_{2}CHO$} & & \citet{1974JMoSp..49..251H} \\
                                   & & & \citet{1980MNRAS.193..713G} \\
                                   & & & \citet{1991RaF....34..213M} \\
                                   & & & \citet{1994RQE...37..155V} \\
                                   & & & \citet{2006JAChS.12812111B} \\
                                   & & & \citet{2009JMoSp.254...28K} \\
                                   & & & \citet{2012AA...548A..71M} \\
                                   \cline{2-4}
                                   & \multirow{2}{*}{Formamide} & \multirow{3}{*}{046512} & \citet{1980MNRAS.193..713G} \\
                                   & \multirow{2}{*}{$\rm NH_{2}^{13}CHO$} & & \citet{2009JMoSp.254...28K} \\
                                   & & & \citet{2012AA...548A..71M} \\
                                   \cline{2-4}
                                   & \multirow{4.5}{*}{Acetamide} & \multirow{5}{*}{059518} & \citet{1987JMoSp.122..408K} \\
                                   & \multirow{4.5}{*}{$\rm CH_{3}C(O)NH_{2}$} & & \citet{1993ZNatA..48..787H} \\
                                   & & & \citet{2001JMoSp.208..188S} \\
                                   & & & \citet{2002JMoSp.215..144Y} \\
                                   & & & \citet{2004JMoSp.227..115I} \\
                                   \cline{2-4}
                                   & $syn$-Glycolamide & \multirow{2}{*}{075517} & \multirow{2}{*}{\citet{2020A&A...639A.135S}} \\
                                   & $syn$-$\rm HOCH_{2}C(O)NH_{2}$ & & \\
\hline
\end{tabular}
\end{table}

\section{The full source model of G358.93 MM1}
\label{app:B}

To make sure that the spectral models are reliable, we added full source model fitting. All species in G358.93 MM1 have been included. The full source model is shown with a solid blue line in Fig. B.1. A complete list of all detected species is provided in Table B.1, together with the LTE parameters adopted for the full source model, including excitation temperature ($T_{\rm ex}$), column density ($N_{\rm t}$), line width (FWHM), and velocity offset ($V_{\rm off}$). For species other than GA, AA, and FA, these parameters were adjusted manually to best reproduce the observed spectrum. Fig. B.1 and Table B.1 are available exclusively on \href{https://doi.org/10.5281/zenodo.19394238}{Zenodo}.

\section{The rotational diagram analysis for GA}
\label{app:C}

\begin{figure}[!htbp]
    \centering
    \includegraphics[width=0.3\textwidth]{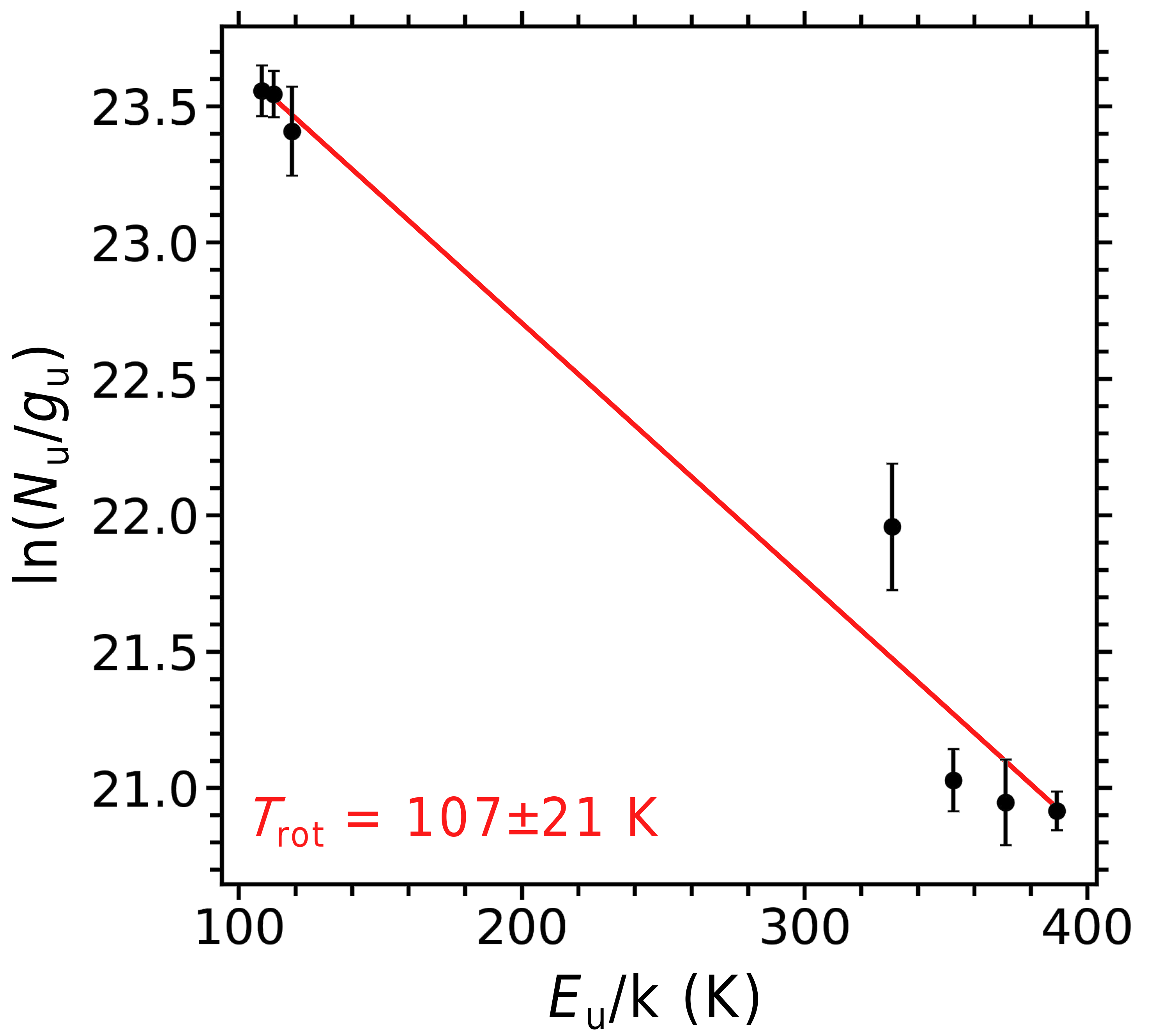}
    \caption{Rotational diagram of GA. Black points represent the observed data, while red lines indicate the best-fit results. The derived rotational temperature ($T_{\rm rot}$) is shown in the lower left corner, with an assumed uncertainty of 20\%.
    \label{fig:C1}} 
\end{figure}

To verify the optimality of the fitting results from CLASS/ADJUST and the validity of the LTE assumption, we have made a rotational diagram analysis (Fig.~\ref{fig:C1}) using seven unblended transitions of GA. The rotational temperature and column density of a given species can be derived using the rotational diagram method, provided that multiple transitions covering distinct upper-level energies are available \citep[e.g.,][]{goldsmith1999, liu2002formic}. Assuming LTE and optically thin emission, the level populations across different transitions can be characterized by a single rotational temperature. Additionally, under the assumption that the molecular emission completely fills the telescope beam, the beam-averaged column density and rotation temperature can be described by the following expression \citep{goldsmith1999, remijan2003, qin2010high}:
\begin{equation}
\centering
\ln\left(\frac{N_u}{g_{up}}\right) = 
\ln\left(\frac{N_t}{Q(T_{\text{rot}})}\right) - \frac{E_{up}}{k T_{\text{rot}}} = 
\ln\left(\frac{8\pi k \nu^2}{h c^3 A_{ul} g_{up}}\right) \int T_{\text{MB}}\, dv,
\label{eq:equation1}
\end{equation}
where $N_{\rm u}$ is the column density of the upper energy level, $g_{\rm up}$ is upper level degeneracy, $N_{\rm t}$ is the total beam-averaged column density, $Q_{\rm rot}$ is the rotational partition function, $E_{\rm up}$ is the upper level energy in K, $k$ is the Boltzmann constant, $T_{\rm rot}$ is the rotation temperature, $v$ is the transition frequency, $h$ is the Planck constant, $c$ is the speed of light, $A_{\rm ul}$ is the Einstein emission coefficient, $\int{T_{\rm MB}}dv$ is the integrated intensity of the specific transition.

The rotational diagram yields a rotational temperature of $107\pm21$ K. Considering the errors, the result agrees reasonably well with the value of $102\pm20$ K derived by the CLASS/ADJUST.

\section{STATCONT-based continuum subtraction and LTE modeling of GA}
\label{app:D}

To address the potential impact of continuum subtraction on the identification of weak spectral features, we repeated the continuum subtraction using STATCONT \citep{Sanchez18}, an automated tool designed to determine the continuum level in line-rich spectra without requiring manual selection of line-free channels. This analysis was performed to verify that our conclusions are not dependent on the specific continuum-subtraction method employed.

Fig. D.1 presents a comparison between the spectra obtained with the CASA-based continuum subtraction and the STATCONT-based method, and can be found exclusively on \href{https://doi.org/10.5281/zenodo.19394238}{Zenodo}. The two spectra are nearly identical across the observed frequency ranges, demonstrating that our manual continuum subtraction procedure does not introduce systematic biases and that the identification of weak spectral features is robust against the choice of continuum-subtraction method.

Using the STATCONT-processed data, we re-identified the candidate GA transitions and performed LTE modeling following the same methodology described in the main text. Fig. D.2 shows the unblended or only slightly blended emission lines (marked with red asterisks) of GA toward G358.93 MM1 identified in the STATCONT-based spectra, and is also available exclusively on \href{https://doi.org/10.5281/zenodo.19394238}{Zenodo}. Compared with the CASA-based analysis, the rms level in the STATCONT-based spectra is slightly higher ($\sim$1.65 K), resulting in five lines detected above 3$\sigma$ instead of six. Nevertheless, the same seven unblended candidate GA lines are recovered, and the derived column density and excitation temperature—$N = (3.7\pm0.7)\times10^{14}$ cm$^{-2}$ and $T_{\rm ex} = 98\pm20$ K—are consistent with those obtained from the CASA-based analysis within the adopted uncertainties. Notably, the column density derived from the STATCONT-based analysis is approximately 30\% lower than that from the CASA-based analysis ($5.3\times10^{14}$ cm$^{-2}$). This comparison shows that, although the derived column density is somewhat method-dependent, our conclusions regarding the tentative identification of GA is not driven by the particular continuum-subtraction approach adopted.

\section{$\rm H_{2}$ column density from dust continuum}
\label{app:E}

We estimated the molecular hydrogen column density, $N \rm _{H_{2}}$, from the 1 mm dust continuum emission under the standard assumption of optically thin thermal dust emission. The beam-averaged column density was calculated using the following relation \citep[e.g.,][]{Mar11, Bon19}:

\begin{equation}
\centering
N_{{\rm H}_{2}} = \frac{{S_{\rm \nu}}{R_{\rm gd}}}{{\mu}{m_{\rm H}}{\Omega}{{\kappa}_{\nu}}{B_{\nu}}({T_{\rm dust}})},
\label{eq:equation1}
\end{equation}
where $S_{\rm \nu}$ is the integrated continuum flux density within the extraction beam, $R_{\rm gd}$ = 100 is the gas-to-dust mass ratio, $\mu$ = 2.8 is the mean molecular weight per $\rm H_{2}$ molecule \citep{Kau08}, $m_{\rm H}$ is the mass of hydrogen atom, $\Omega$ is the beam solid angle, ${{\kappa}_{\nu}}$ is the dust opacity, and ${{B_{\nu}}({T_{\rm dust}})}$ is the Planck function evaluated at the dust temperature $T_{\rm dust}$. We adopt ${{\kappa}_{\nu}}$ = 2 cm$^{2}$ g$^{-1}$ at ${\rm \lambda}$${\sim}$1 mm \citep{Bon19} and $T_{\rm dust}$ = 150 K, following \citet{2020NatAs...4.1170C}. With these assumptions, we derive $N \rm _{H_{2}}$ = (7.82$\pm$1.56)$\times$10$^{24}$ cm$^{-2}$ for G358.93 MM1, where a conservative uncertainty of 20\% is assumed. Molecular abundances relative to $\rm H_{2}$ were then calculated as $\chi = N_{\rm t}/N \rm _{H_{2}}$, and the resulting $\chi$ values are reported in Table \ref{tab:1}.

\end{appendix}

\end{document}